\documentclass[singlecolumn, notitlepage,          % Format : preprint, twocolumn
               showpacs,            % Pacs : showpacs, noshowpacs
               preprintnumbers,     % Preprint: preprintnumbers,
               aps,                 % Society: ...
               prd,          	    % Journal Style : pra, prb, prc, prd, pre,
               a4paper,   11pt,          % Size : a4paper, ...
               superscriptaddress,      % Affiliation (Title) : groupedaddress,
               nofootinbib,         % Footnote: footinbib, nofootinbib
               tightenlines,        % Remove additional spaces in a line
               floats,floatfix      % Floating pictures and tables
               ]{revtex4-1}
\usepackage{graphicx}  % needed for figures
\usepackage{dcolumn}   %needed for some tables
\usepackage{bm}        % for math
\usepackage{amsmath,amssymb}
\usepackage{float}
\usepackage{makecell}
\usepackage[thinlines]{easytable}
\usepackage{array,booktabs}
\usepackage{longtable}
\usepackage{lipsum}
\usepackage{array,makecell}
\newcolumntype{P}[1]{>{\centering\arraybackslash}p{#1}}

\begin{document}

\title{Second Order Perturbative Effects on the Acoustic Geometry }% Force line breaks with \\

\author{Nandan Roy}  
 \email{nandan@fisica.ugto.mx}
 \affiliation{%
Departamento de F\'isica, DCI, Campus Le\'on, Universidad de
Guanajuato, 37150, Le\'on, Guanajuato, M\'exico.} 

\author{Shivani Singh}%
\affiliation{Indian Institute of Technology, Hyderabad, Telengana 502285, India
}%
\email{ph13m1013@iith.ac.in}

\author{Sankhasubhra Nag}%
\affiliation{ Sarojini Naidu College for Women, Kolkata 700028, India.
}%
 \email{sankha@sncwgs.ac.in}

\author{Tapas K. Das}%
\affiliation{ Harish-Chandra Research Institute, Allahabad, Jhunsi 211019, India
}%
 \email{tapas@hri.res.in}

\begin{abstract}
In this work, we have studied the effect of higher order perturbations, particularly the second order in details, on the sonic horizon. We have considered two different schemes of perturbations which are velocity potential perturbation and mass acceleration rate perturbation. These two schemes give us qualitatively similar behaviour.  We have found that the analogue gravity formalism also holds for the higher order perturbations.

\end{abstract}

\pacs{04.25.-g, 04.30.Db, 97.60.Jd, 11.10.Gh}

\keywords{Analogue gravity, black hole accretion, higher order perturbation}

\maketitle

%\tableofcontents
\section{Introduction}

In 1981 Unruh \cite{unruh1981experimental} developed a way of mapping certain aspects of Black Hole physics into the theory of transonic fluid flows. He showed that for an inviscid and irrational flow, the linear perturbation of the velocity potential lives in a curved space-time though the underlying physical flow of the fluid is Newtonian. The perturbations obey the Klein Gordon equation for a massless scalar field in curved space time. Physically, for a spherically symmetric, transonic and converging flow one can find a dumb hole. The radius at which the bulk velocity crosses the sound velocity works as the event horizon of the linear perturbations or the sound waves. Mathematically it is possible to map these behaviour of the dumb hole to the Schwarzschild black holes. In \cite{visser1998acoustic}, Visser discussed more details on the analogy between the black holes of Einstein gravity and supersonic fluid flows.

    This analogue gravity phenomenon can also be studied in astrophysical accretion.  For a transonic accretion in which the bulk velocity of the fluid gradually increases from the subsonic to supersonic the sound wave trapped inside the supersonic region will not be able to cross the sonic horizon. As for example of the transonic accretion, one can see, Bondi accretion \cite{bondi1952spherically} and sub-Keplerian disk accretion \citet{chakrabarti2001model} at the center of our galaxy. Mass accretion rate is a very important physical quantity in the accretion phenomena. The linear perturbation of the mass accretion rate can also give rise to analogue gravity phenomena\cite{nag2012role}. Recently it has been also shown that perturbation of Bernoulli constant also leads to analogue gravity \cite{datta2016acoustic}. In the General Relativistic framework, several studies of the astrophysical accretion and formation of the acoustic black hole have been done. Bilic in \cite{bilic1999relativistic} shows that perturbation of the velocity potential in the curved spacetime also has analogue gravity effect. Similarly, \cite{ananda2015acoustic,bollimpalli2017perturbation} discuss the perturbation of mass accretion rate and stability of the system.
    
    In this work, we have tried to find out the effect of second-order perturbations on the acoustic geometry. We have adopted both velocity potential and mass accretion rate perturbation schemes. These two different schemes give us qualitatively same results. It is found that the higher order perturbations also see the same acoustic metric as seen by the first order but the dynamics of the higher order is effected by the lower order perturbations. The lower order perturbations appear as the source in the wave equation of the higher order perturbations. 
    
    The paper is structured in the following way; section II starts with a very basic description of the fluid dynamics, in section III we discuss the mathematical derivation of the wave equations for the fluctuations using velocity potential perturbation and mass accretion rate perturbations, section IV is a general discussion on the result obtained and we conclude the paper with the concluding remarks in section V.

\section{Basic description}
      	 Let us consider a nonrelativistic fluid in which the fluid velocity is much less than light velocity. The momentum conservation and mass conservation of the system follow the Newtonian dynamics. The continuity equation of the fluid is given by 
      	 
      	 \begin{equation} \label{conti}
      	 \partial_t \rho + \triangledown . (\rho \vec{v}) = 0,
      	 \end{equation} 
   where $\rho$ and   $\vec{v}$ are the  mass density and velocity of the fluid.	 
      	The Euler's Equation for an inviscid flow can be written as 
      	 
      	 \begin{equation}
      	 \rho \dfrac{d\vec{v}}{dt} \equiv \rho[\partial_t \vec{v} + (\vec{v} \cdot \triangledown)\vec{v}] = \vec{F}.
      	 \end{equation}
      	 
      	 Here,
      	 
      	 \begin{equation}
      	 \vec{F} = - \triangledown p - \rho \triangledown \Phi,
      	 \end{equation}
      	 
      	 and $\Phi$ denotes the Newtonian gravitational potential and external driving force.

      	For an irrotational flow, the velocity of the fluid can be written in terms of velocity potential $\psi$ such that $\vec{v} = -\triangledown \psi$, at least locally. If one considers the fluid to be barotropic, it becomes possible to define the specific enthalpy as,
      	 
      	 \begin{equation}
      	 h(p) = \int_{0}^{p} \dfrac{dp'}{\rho (p')} ~~; 
      	 \end{equation} 
      	 
      	 so that  
      	 
      	\begin{equation} \label{enthalpy}
      	\triangledown h = \dfrac{1}{\rho} \triangledown p.
      	\end{equation}
      	
      	Using the irrotationality condition, $\triangledown \times v = 0$ one can manipulate the Euler's equation to be written in the following form,
      	
      	\begin{equation} \label{euler}
      	- \partial_t \psi + h + \dfrac{1}{2} (\triangledown \psi)^2 +  \Phi = 0 
      	\end{equation}
      	
      	This is Bernoulli's equation for the irrotational and inviscid fluid flow in the presence of external driving forces.
    
   \section{Fluctuations}
   
Physically, sound waves in a moving fluid are always dragged by it. If the flow is a transonic flow which means the speed of the fluid gradually increase and becomes supersonic from subsonic then the sound waves cannot swim upstream and become trapped inside the supersonic region. This region is known as the dumb hole, an inescapable region for sound waves. This is analogs to the notion of a black hole. This analogy is not only verbal it is also mathematical. 

Mathematically, the sound waves are defined as the linearized fluctuations in the hydrodynamic quantities. Though the underlying fluid dynamics is Newtonian it can be shown that these fluctuations live in a (3+1)-dimensional curved Lorentzian space-time. To study the behavior of the fluctuations there are two different perturbation schemes. One can perturb the velocity potential and also the mass accretion rate. For linearized perturbations, these two schemes qualitatively gave the same result. In this work, our goal is to consider nonlinear fluctuations and analyze their effect on acoustic geometry using these two different perturbation schemes. We are also interested to check if this equivalence between these two perturbation schemes is, in general, true for higher order perturbations also.
   
   \subsection{Velocity Potential Perturbation}
Following is the description of the perturbation scheme we are going to use,

\begin{subequations}
\begin{eqnarray}
\rho = \rho _0 + \epsilon \rho _1 + \epsilon ^2 \rho _2 + O(\epsilon ^3), \\
p = p_0 + \epsilon p_1 + \epsilon ^2 p_2 + O(\epsilon ^3),\\
\psi = \psi _0 + \epsilon \psi _1 + \epsilon ^2 \psi _2 + O(\epsilon ^3)
   \end{eqnarray}
   \end{subequations}

We have also considered that $\rho_2 << \rho_1, p_2 << p_1 $ and $\psi_2 << \psi_1 $ which physically implies that the second modes of perturbation are very small compared to the amplitude of the first mode of the perturbation.
One more consideration is that the origin of the fluctuations are independent and they do not add up to give a single resultant perturbation.

From the equation (\ref{enthalpy}), we can see the specific enthalpy $h(p)$ is function of pressure so it will be also perturbed in the same fashion as $h = h_0 + \epsilon h_1 + \epsilon ^2 h_2 +O(\epsilon ^3)$. 
   
   Substitution of the perturbed values in the modified Euler's equation (\ref{euler}) and equating the coefficient of different powers of $\epsilon$ :
   
   \begin{equation} \label{psi_1}
   -\partial_t \psi _1 + h_1 + (\triangledown \psi _0)(\triangledown \psi _1) = 0,
   \end{equation}
   
   and
   
   \begin{equation} \label{psi_2}
   -\partial_t \psi _2 + h_2 + \dfrac{1}{2}(\triangledown \psi _1)^2 + (\triangledown \psi _0)(\triangledown \psi _2) = 0.
   \end{equation}
   
   Similarly, on substituting the perturbed values in continuity equation (\ref{conti}) and equating the coefficient of different powers of $\epsilon$, we get 
   
   \begin{equation} \label{rho_1}
   \partial_t \rho_1 + \triangledown . (\rho_0 \vec{v_1} + \rho_1 \vec{v_0}) = 0
   \end{equation}
   
   and 
   
   \begin{equation} \label{rho_2}
    \partial_t \rho_2 + \triangledown . (\rho_0 \vec{v_2} + \rho_1 \vec{v_1} + \rho_2 \vec{v_0}) = 0
   \end{equation}

We have also found out different coefficient of perturbations of $h(p)$ using it's definition and Taylor expansion around the background variables. Following are the different coefficient of $h(p)$, 
   
  \begin{eqnarray} \label{h}
  &&  h_0 = \dfrac{c_{s0}^2}{(\gamma-1)} \ , \nonumber\\
  &&  h_1 = c_{s0}^2 \dfrac{\rho_1}{\rho_0} \ , \nonumber \\
  &&  h_2 = c_{s0}^2 (\dfrac{\rho_2}{\rho_0} + \dfrac{(\gamma -2)}{2} \dfrac{\rho_1 ^2}{\rho_0 ^2}).
  \nonumber \\
  \end{eqnarray}
  
  In the modified Euler equation (\ref{euler}), substituting the value of $h_1$ from equation (\ref{h}) and finding the value of $\rho_1$,
  
  \begin{equation}
  \rho_1 = \dfrac{\rho_0}{c_{s0}^2} [\partial _t \psi_1 + v_0 \triangledown \psi_1]
  \end{equation}
   
   and on differentiation of the above equation with respect to t gives,
   
   \begin{equation}
   \partial_t \rho_1 = \dfrac{\rho_0}{c_{s0}^2} [\partial _t ^2 \psi_1 + v_0 \partial_t(\triangledown \psi_1)] .
   \end{equation}
   
   Now substituting this value in equation (\ref{rho_1}) and rearranging, the wave equation for the first order perturbation can be written in a compact form as,
   
   \begin{equation}\label{velo1}
   \partial_{\mu}(\eta^{\mu \nu} \partial_{\nu} \psi_1) = 0
   \end{equation}
   
   here, 
   
   \begin{eqnarray} \label{eta_velo}
   && \eta^{tt} = \dfrac{\rho_0}{c_{s0}^2} ; \nonumber \\
   && \eta^{tr} \equiv h^{rt} = \dfrac{v_0 \rho_0}{c_{s0}^2} ; \nonumber \\
   && \eta^{rr} = \dfrac{\rho_0}{c_{s0}^2} [v_0 ^2 - c_{s0}^2]
   \end{eqnarray}
   
   Similarly for second order perturbation, in equation (\ref{psi_2}) substituting the value of $h_2$ from equation (\ref{h}) and we obtain an expression for $\rho_2$
   
   \begin{equation}
   \rho_2 = - \dfrac{(\gamma -2)}{2} \dfrac{\rho_1 ^2}{\rho_0} + \dfrac{\rho_0}{c_{s0}^2}[\partial_t \psi_2 + v_0 \triangledown \psi_2 + \dfrac{v_1}{2} \triangledown \psi_1]
   \end{equation}
   
   Now substituting this value of $\rho_2$ and $\rho_1$ in equation (\ref{rho_2}), we get an coupled wave equation for second order perturbation as,
   
   \begin{equation} \label{velo2}
   \partial_{\mu} (\eta^{\mu \nu}\partial_{\nu}\psi _2) - \partial_{\mu} (g^{\mu \nu}\partial_{\nu}\psi _1) = 0 ,
   \end{equation}
   
   where the components of $\eta^{\mu \nu}$ are the same as in the first order perturbations (\ref{eta_velo}). The component of the $g^{\mu \nu}$ are following,

   \begin{widetext}
   \begin{eqnarray}
   && g^{tt} = \dfrac{(\gamma-2)}{2}\dfrac{\rho_1}{c_{s0}^2} ; \nonumber \\
   && g^{tr} \equiv g^{rt} = \dfrac{(\gamma-2)}{2}\dfrac{ \rho_1}{c_{s0}^2} [v_0 - \dfrac{\rho_0}{(\gamma -2)\rho_1} v_1] ; \nonumber \\
   && g^{rr} = \dfrac{(\gamma-2)}{2}\dfrac{\rho_1}{c_{s0}^2} [(v_0 ^2 - c_{s0}^2) + \dfrac{(\gamma-1)}{(\gamma-2)} c_{s0}^2 - \dfrac{\rho_0}{(\gamma-2)\rho_1} 2 v_0 v_1].
   \end{eqnarray}
   \end{widetext}

\subsection{Mass Accretion Rate}
    
The mass accretion rate is defined as,
    
    \begin{equation} \label{mass_rate}
    f = \rho v r^2.
    \end{equation}
    
    We will study non-linearized fluctuations around background values $(\rho _0 , v_0 , f_0)$, set as, $ \rho = \rho_0 + \epsilon \rho _1 + \epsilon ^2 \rho _2 + O(\epsilon ^3)$ , $v = v_0 + \epsilon v_1 + \epsilon ^2 v_2 + O(\epsilon ^3)$ , and $f = f_0 + \epsilon f_1 + \epsilon ^2 f_2 + O(\epsilon ^3)$. Hence equation (\ref{mass_rate}) will be, 
    
    \begin{equation}
    f_0 + \epsilon f_1 + \epsilon ^2 f_2 + O(\epsilon ^3) = [\rho_0 + \epsilon \rho _1 + \epsilon ^2 \rho _2 + O(\epsilon ^3)][v_0 + \epsilon v_1 + \epsilon ^2 v_2 + O(\epsilon ^3)] r^2.
    \end{equation}
   
    Then equating the coefficient of different powers of $\epsilon$ will give, 
    
    \begin{equation}
    f_0 = \rho_0 v_0 r^2,
    \end{equation}
    
    \begin{equation} \label{f1}
    f_1 = (\rho_0 v_1 + \rho_1 v_0) r^2,
    \end{equation}
        
    \begin{equation} \label{f2}
    f_2 = (\rho_0 v_2 + \rho_1 v_1 + \rho_2 v_0) r^2.
    \end{equation}
    
   We consider the flow to be radial, therefore continuity equation reduces to,
   
   \begin{eqnarray}
   && \partial_t \rho + \dfrac{1}{r^2} \partial_r (r^2 \rho \vec{v}) = 0 , \nonumber \\
   && \partial_t \rho = - \dfrac{1}{r^2} \partial_r f 
   \end{eqnarray}
   
   On substituting and equating the coefficients of different powers of $\epsilon$ we get,
   
   \begin{eqnarray}
   &&\partial_t \rho_1 = - \dfrac{1}{r^2} \partial_r f_1 , \\
   &&\partial_t \rho_2 = - \dfrac{1}{r^2} \partial_r f_2.
   \end{eqnarray}
    
   Differentiation of the equation (\ref{f1}) with respect to 't' will give the value of $\partial_t \rho_1$
   
   \begin{equation}
   \partial_t \rho_1 = [\dfrac{1}{v_0 r^2} \partial_t f_1 - \dfrac{\rho_0}{v_0} \partial_t v_1].
   \end{equation} 
   
   Substitution of the value of $\partial_t \rho_1$ in equation (26)  gives,
   
   \begin{equation}
   \partial_t v_1 = \dfrac{v_0}{f_0} [\partial_t f_1 + v_0 \partial_r f_1].
   \end{equation}
    
    Similarly, differentiating equation  (\ref{f2}) with respect to 't' one will get the value of $\partial_t \rho_2$
    
    \begin{equation}
    \partial_t \rho_2 = [\dfrac{1}{v_0 r^2} \partial_t f_2 - \dfrac{\rho_0}{v_0} \partial_t v_2 + \dfrac{v_1}{r^2 v_0} \partial_r f_1 - \dfrac{\rho_1}{f_0} \partial_t f_1 - \dfrac{\rho_1 v_0}{f_0} \partial_r f_1].
    \end{equation}
     
    On substituting this value in equation (27), we get
    
    \begin{equation}
    \partial_t v_2 = \dfrac{v_0}{f_0} [\partial_t f_2 + v_0 \partial_r f_2] - \dfrac{\rho_1 v_0}{\rho_0 f_0}[\partial_t f_1 + (v_0 - \dfrac{\rho_0}{\rho_1} v_1) \partial_r f_1].
    \end{equation}
     
    Since it is not a steady state solution, fluctuation in $c_{s} ^2$ is given by,
    \begin{widetext}
    \begin{equation}
    c_s ^2 = c_{s0}^2 + \epsilon (\gamma -1) \dfrac{\rho_1}{\rho_0} c_{s0} ^2 + \epsilon^2 [(\gamma -1) \dfrac{\rho_2}{\rho_0} c_{s0} ^2 + \dfrac{1}{2} (\gamma -1) (\gamma -2) \dfrac{\rho_1 ^2}{\rho_0 ^2} c_{s0}^2] + O(\epsilon ^3).
    \end{equation}
    \end{widetext}
     
    Hence,
     \begin{widetext}
    \begin{equation}
    \dfrac{c_{s}^2}{\rho} = \dfrac{c_{s0}^2}{\rho_0} + \epsilon [(\gamma - 2) \dfrac{\rho_1}{\rho_0 ^2} c_{s0}^2] + \epsilon^2 [(\gamma - 2) \dfrac{\rho_2}{\rho_0 ^2} c_{s0}^2 + (\gamma -2)(\gamma -3)\dfrac{\rho_1^2}{\rho_0 ^3} c_{s0}^2] + O(\epsilon ^3).
    \end{equation}
    \end{widetext}
    
    Substituting $\partial_r p = c_{s}^2 \partial_r \rho$ and taking the time derivative of equation (3), we get
    
    \begin{equation}
    \partial_t ^2 v + \partial_r[v \partial_t v + \dfrac{c_s ^2}{\rho} \partial_t \rho] = 0.
    \end{equation}
    
    After putting the perturbed value in the above equation and  
    equating the coefficient of different power of $\epsilon$, we get
    
    \begin{equation}
    \partial_t ^2 v_1 + \partial_r [v_0 \partial_t v_1 + \dfrac{c_{s0}^2}{\rho_0} \partial_t \rho_1] = 0,
    \end{equation}
     
     \begin{equation}
     \partial_t ^2 v_2 + \partial_r[(v_1 \partial_t v_1 + v_0 \partial_t v_2) + (\dfrac{c_{s0}^2}{\rho_0} \partial_t + (\gamma - 2) \dfrac{\rho_1 c_{s0}^2}{\rho_0 ^2} \partial_t \rho_1)] = 0.
     \end{equation}
     
    Hence, for First Order Perturbation :
    
    Substituting the value of $\partial_t v_1$ and $\partial_t \rho_1$ from equation (28) and (29) in equation (35), we get
    
    \begin{equation}
    \partial_{\mu} (h^{\mu \nu} \partial_{\nu} f_1) = 0,
    \end{equation} 
    
    where,
    
     \begin{eqnarray} \label{h.mass}
     && h^{tt} = \dfrac{v_0}{f_0} ; \nonumber \\
     && h^{tr} \equiv h^{rt} = \dfrac{v_0 ^2}{f_0} ; \nonumber \\
     && h^{rr} = \dfrac{v_0}{f_0} [v_0 ^2 - c_{s0}^2].
     \end{eqnarray}
     
For second order perturbation substituting the value of $\partial_t v_1$, $\partial_t v_2$, $\partial_t \rho_2$ and $\partial_t \rho_1$ from equations (28 -31) in (36) we get
    
    \begin{equation}
    \partial_{\mu} (h^{\mu \nu}\partial_{\nu}f_2) - \partial_{\mu} (g^{\mu \nu}\partial_{\nu}f_1) = 0 ,
    \end{equation}
    
    where $h^{\mu \nu}$ are the same as the equation (38) and the component of $g^{\mu \nu}$ is given below,
    
    \begin{eqnarray}
    && g^{tt} = \dfrac{\rho_1}{\rho_0} \dfrac{v_0}{f_0}; \nonumber \\
    && g^{tr} \equiv g^{rt} = \dfrac{\rho_1}{\rho_0} \dfrac{v_0}{f_0} [v_0 - \dfrac{\rho_0}{\rho_1} v_1] ; \nonumber \\
    && g^{rr} = \dfrac{\rho_1}{\rho_0} \dfrac{v_0}{f_0} [(v_0 ^2 - c_{s0}^2) + (\gamma-1) c_{s0}^2 - \dfrac{\rho_0}{\rho_1} 2 v_0 v_1].
    \end{eqnarray}

    \section{General discussion on the result obtained}
    
    In this section, we discuss the results we obtained due to the inclusion of second-order perturbation. For both velocity potential perturbation and mass accretion rate perturbation the qualitative results are same. We get back the same good old equation for the first order perturbations. For velocity potential $\partial_{\mu}(\eta^{\mu \nu} \partial_{\nu} \psi_1) = 0$ and for mass accretion rate $ \partial_{\mu} (h^{\mu \nu} \partial_{\nu} f_1) = 0$, where the exact form of different component of $\eta^{\mu \nu}$ and $h^{\mu \nu}$ are given in the equations (\ref{eta_velo}) and (\ref{h.mass}). These  wave equations for the first order perturbations can be written in the form of Lorentzian manifold the curved space scalar d’Alembertian as,
    
    \begin{equation}
    \triangle \phi = \frac{1}{\sqrt[]{-g}} \partial_{\alpha} \left(\sqrt[]{-g} g^{\alpha \beta} \partial_{\beta} \phi \right) = 0,
    \end{equation}
   
    provided that the identity $\sqrt[]{-g} g^{\mu \nu} = \eta^{\mu \nu} = h^{\mu \nu} $ holds. By further manipulation one can write down the acoustic line element as 
   
    \begin{equation}
     ds^2 \equiv dx^\alpha dx^\beta = \frac{\rho_0}{c} \left( -c^2 dt^2 +  (dx^i - v^i_0 dt) \delta_{ij}(dx^j - v^j_0 dt) \right).
    \end{equation}
   This line element can be mapped to the Painleve–Gullstrand line element which is an unusual representation of the Schwarzschild geometry. It is a very interesting phenomenon in which though the underlying fluid flow is Newtonian, non-relativistic and takes places in a flat spacetime, the first order perturbations live in a curved spacetime which can be mapped to Schwarzschild geometry. For the second order perturbations, the wave equations are coupled and are given as
   
   \begin{equation}
    \partial_{\mu} (\eta^{\mu \nu}\partial_{\nu}\psi _2) - \partial_{\mu} (g^{\mu \nu}\partial_{\nu}\psi _1) = 0 ,
    \end{equation}
    
    \begin{equation}
   \partial_{\mu} (h^{\mu \nu}\partial_{\nu}f_2) - \partial_{\mu} (g^{\mu \nu}\partial_{\nu}f_1) = 0. 
   \end{equation}

It is interesting to note that the background acoustic metric ($\eta^{\mu \nu}, h^{\mu \nu}$) is with the second order perturbations and the first order perturbations work as a source term to the wave equation of the second order wave equations. So the second order perturbations see the same acoustic metric but its dynamics is not free as it is influenced by the first order perturbations. Travelling wave and standing wave solution of the first order perturbations and the corresponding stability criteria is found out in \cite{bollimpalli2017perturbation}. We leave finding out of the solution of the second order perturbation as our future research topic. This method can be extended to more higher order perturbations. We have checked it that always the higher order perturbations see the background acoustic geometry which is created by the first order perturbations and all lower perturbations work as the source.

    \section{Concluding remarks} In this work, we have checked the effect of higher order perturbations on the acoustic geometry. We have considered two different types of perturbation schemes namely, velocity potential perturbation and mass accretion rate perturbation. These two perturbations give us the qualitatively same result. In this work our central assumption is, the mode of the perturbations are independent of each other and they do not add up to give a single perturbation. The acoustic geometry is the effect of the first order perturbation. The second order perturbations experience the same metric but its evolution is affected by the presence of the first order perturbations.  
    
    The present work reports the preliminary results obtained while performing a more extended research project on the non-linear higher order perturbation in accretion Astrophysics. Such extended work is in progress~\cite{Roy} and will be reported elsewhere.
    
    \section{Acknowledgements}
    The authors would like to deeply express their gratitude to J. K. Bhattacharjee for very useful and stimulating discussions. SS visited HRI twice under the scheme of visiting students research programme of HRI. Visits of SN have been supported by the 12th plan budget of the Cosmology and high energy Astrophysics project fund. NR was a post doctoral fellow of TKD at HRI when the present work was performed.
    
%\bibliography{perturbation} 

\end{document}